\documentclass[prl,aps,reprint,citeautoscript,amsmath,amssymb,superscriptaddress]{revtex4-1}
\usepackage{graphicx}
\usepackage{dcolumn}
\usepackage{bm}
\usepackage{amsmath}
\usepackage{soul} 
\DeclareGraphicsExtensions{.pdf,.eps,.png,.jpg,.mps}

\usepackage{soul} 
\usepackage[utf8]{inputenc} 
\usepackage{hyperref} 
\usepackage{xcolor} 

\usepackage{fixfoot}

\DeclareFixedFootnote{\repnote}{See Supplemental Material at [URL will be inserted by publisher] for a more complete description of our magnetized 3OTB Hamiltonian, a detailed characterization of the system when vacancies are present, spin current calculations without/with vacancies, and mean field estimations of electron-electron interaction effects, which includes Refs.[algo1-algo2]}

\begin{document}

\title{Tunable spin-polarized edge currents in proximitized transition metal dichalcogenides}

\author{Natalia Cortés}
\email{natalia.cortesm@usm.cl}
\affiliation{Departamento de Física, Universidad Técnica Federico Santa María, Casilla 110V, Valparaíso, Chile}
\affiliation{Department of Physics and Astronomy, and Nanoscale and Quantum Phenomena Institute, Ohio University, Athens, Ohio 45701–2979, USA}
\author{O. Ávalos-Ovando}
\affiliation{Department of Physics and Astronomy, and Nanoscale and Quantum Phenomena Institute, Ohio University, Athens, Ohio 45701–2979, USA}
\author{L. Rosales}
\affiliation{Departamento de Física, Universidad Técnica Federico Santa María, Casilla 110V, Valparaíso, Chile}
\author{P. A. Orellana}
\affiliation{Departamento de Física, Universidad Técnica Federico Santa María, Casilla 110V, Valparaíso, Chile}
\author{S. E. Ulloa}
\affiliation{Department of Physics and Astronomy, and Nanoscale and Quantum Phenomena Institute, Ohio University, Athens, Ohio 45701–2979, USA}

\date{\today}

\begin{abstract}
We explore proximity-induced ferromagnetism on transition metal dichalcogenides (TMDs), focusing on molybdenum ditelluride (MoTe$_2$) ribbons with zigzag edges, deposited on ferromagnetic europium oxide (EuO). A tight-binding model incorporates exchange and Rashba fields induced by proximity to EuO or similar substrates.  For in-gap Fermi levels, electronic modes in the nanoribbon are localized along the edges, acting as one-dimensional (1D) conducting channels with tunable spin-polarized currents. TMDs on magnetic substrates can become very useful in spintronics, providing versatile platforms to study proximity effects and electronic interactions in complex 1D systems.
\end{abstract}


\maketitle
\date{Today}

\emph{Introduction.} The successful combination of two-dimensional (2D) materials with magnetic insulator substrates \cite{swartz2012integration,wei2016strong,leutenantsmeyer2016proximity,Zhao2017,Zhong2017van} has opened interesting possibilities to exploit material properties and create novel functionalities \cite{vzutic2004spintronics,pesin2012spintronics}. Interactions between spins in a non-magnetic material and those from a ferromagnetic (FM) or antiferromagnetic crystal in close proximity have expanded spintronics research \cite{amamou2018}.  The interactions may be due to non-vanishing wave-function overlap of localized moments in the magnetic crystal with electrons in a 2D layer \cite{semenov2007spin, haugen2008spin,li2018large}. A spin splitting of 5 meV was predicted for monolayer graphene deposited on FM europium oxide (EuO) \cite{haugen2008spin,Yu2015APL}, motivating the successful epitaxial growth of EuO on graphene \cite{swartz2012integration}.
Experiments with different FM substrates have reported magnetic exchange fields (MEF) induced on graphene of $\sim$14 T (when on EuS) \cite{wei2016strong}, and $\sim$0.2 T (on YIG) \cite{leutenantsmeyer2016proximity}. Proximitized interactions clearly allow for the effective control of the spin degree of freedom in 2D materials, a fundamental element in spintronic devices.

Transition metal dichalcogenide (TMD) monolayers built as $MX_{2}$ ($M$ = Mo, W, and $X$ = S, Se, Te) \cite{novoselov2005two,liu2015electronic} exhibit a direct bandgap located at the $K$ and $K'$ points in the Brillouin zone \cite{Mak2010,wang2012electronics}. Spin-orbit coupling (SOC) and intrinsic lack of inversion symmetry cause a sizable spin splitting at the valence band edges \cite{Xiao2012}. TMD valleytronics applications require the lifting of valley degeneracy, which has been achieved only by application of large magnetic fields \cite{2018arXiv180507942Z}, so that
magnetic proximity effects may provide more practical alternatives \cite{Zhao2017,Zhong2017van}. Indeed, proximity effects on TMD monolayers when on a FM insulator substrate \cite{qi2015giant,du2016rashba,liang2017magnetic,Habe2017,xu2018large,li2018large,Seyler2018} are predicted to lift valley degeneracy due to broken time reversal symmetry (TRS) and exchange fields \cite{qi2015giant,Zhang2016,liang2017magnetic}. Experiments found a few meV valley splitting in a WSe$_{2}$ monolayer on a EuS FM substrate \cite{Zhao2017}. When deposited on another FM substrate CrI$_{3}$ \cite{Zhong2017van,Seyler2018}, WSe$_{2}$ was found to exhibit a slightly larger valley splitting ($\simeq 3.5$ meV).
A giant splitting (300 meV) due to the induced MEF was predicted for a MoTe$_{2}$ monolayer on EuO \cite{qi2015giant,Zhang2016}, together with a sizable Rashba field ($\lesssim 100$ meV). As we will see, the competition between Rashba and exchange fields provides an important figure of merit that determines the behavior of such systems.

In this paper we provide the missing piece on how TMD edges are affected by proximitized magnetism. As in graphene, the hexagonal lattice in TMDs allows clean edges to be labeled as zigzag or armchair-terminated, with the first being much more common and stable in the lab \cite{vasu2015clean,Cheng2017nanoletters,Chen2017natcomms,Zhao2017nanoletters,chen2017atomically,huangsitu}. Zigzag-terminated TMD structures reveal rich one-dimensional behavior, such as metallic edge modes \cite{Gibertini2015,rostami2016edge,Chu2014,ridolfi2017electronic,Cheng2017nanoletters}, and helical states that host Majorana bound states at the ends of a ribbon \cite{Chu2014}. Twin boundaries have also been shown to host 1D charge density waves \cite{barja2016charge}. Non-magnetic 1D edge states have also been reported recently in topological superconductors \cite{schnyder2013edge}, graphene superlattices \cite{brown2018edge}, and for graphene on TMDs \cite{frank2018}.

We analyze a zigzag MoTe$_{2}$ ribbon deposited on a FM substrate such as EuO \cite{qi2015giant,Zhang2016}.
The proximity-induced ferromagnetism is incorporated through a real space three-orbital tight-binding (3OTB) Hamiltonian \cite{Liu2013}, that allows us to explore the electronic eigenstates and associated spin-polarized currents in the proximitized MoTe$_{2}$ ribbon. We find edge modes that are spatially confined to the zigzag edges, are strongly spin-polarized, and act as effective 1D conducting channels that carry spin-polarized currents while the bulk is insulating.
The effect of defects in the structure and response is also analyzed; we find spin currents to be robust for moderate vacancy concentrations, before being suppressed in a highly defective system. The strong MEF suppresses antiferromagnetic ground states on the edges that may arise from electronic interactions, favoring the spin polarization that gives rise to the edge currents.   The generic existence of TMD edge states that can be accessed by gating suggests that these hybrid systems could be used as robust tunable spin filters for use in diverse applications, apart from providing interesting systems to explore proximity magnetism, and the role of electronic interactions in 1D systems with complex spin texture. This system complements recent spin current experiments on devices containing EuO and/or 2D materials \cite{Zhang2018PRL,Leutenantsmeyer2018PRL,Xu2018PRL}, bringing TMDs to this exciting area. 


\emph{Model.} To describe the low-energy spectrum of a commensurate FM/TMD heterostructure \cite{qi2015giant,Scharf2017},
we generalize a successful 3OTB model \cite{Liu2013} to include magnetic exchange field effects \cite{supplemental}. The model has relevant lattice symmetries and has been proven to reliably describe TMD ribbons \cite{Chu2014} and flakes in diverse situations \cite{Segarra2016,Avalos2016v2,Pawlowski2018,Alsharari2018}.
\begin{figure}[!h]
\centering
\includegraphics[width=1.00\linewidth]{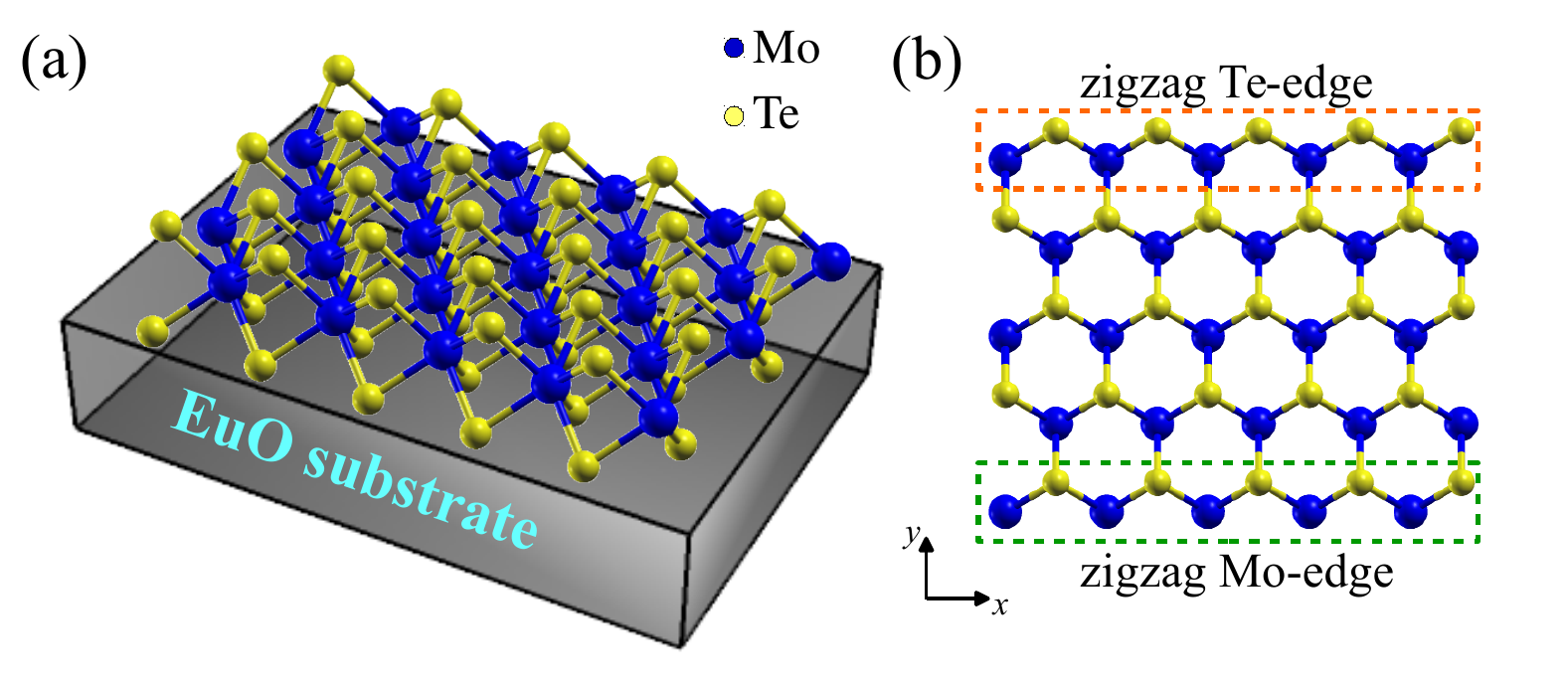}
\caption{(a) Schematic representation of a MoTe$_{2}$ ribbon in proximity to a ferromagnetic substrate such as EuO. (b) Top view of MoTe$_{2}$ ribbon with zigzag Te and Mo-edges. Eu atoms are hidden directly below the Mo atoms (blue spheres).}
\label{fig1}
\end{figure}
The nearly commensuration of the MoTe$_{2}$/EuO structure \cite{qi2015giant,Zhang2016}, as the EuO (111) surface and TMD lattice have only a 2.7\% mismatch, incorporates the substrate effects into the pristine MoTe$_{2}$ as on-site magnetic exchange and Rashba fields, as $\mathcal{H}_{\mathrm{MoTe}_{2}\mathrm{/EuO}}=\mathcal{H}_{\mathrm{MoTe}_{2}}+\mathcal{H}_{\mathrm{ex}}+\mathcal{H}_{\mathrm{R}}$ \cite{supplemental}. Here, $\mathcal{H}_{\text{MoTe}_{2}}$ is the pristine TMD Hamiltonian \cite{Liu2013}. It is written in a basis of relevant transition metal $d$-orbitals, $\big\{ \left|d_{z^{2}},s \right\rangle$, $\left|d_{xy},s \right\rangle$, $\left|d_{x^{2}-y^{2}},s \right\rangle \big\}$, with spin index $s=\uparrow,\downarrow$ \cite{Liu2013}.
The induced MEF is spin diagonal, with blocks $\mathcal{H_\mathrm{ex,\uparrow \uparrow}}=-\mathcal{H_\mathrm{ex,\downarrow \downarrow}}=\mathrm{diag}\{-B_{c},-B_{v},-B_{v}\}$, where $B_{c}=206$ meV and $B_{v}=170$ meV correspond to conduction and valence exchange fields, respectively \cite{qi2015giant}. A Rashba field also arises from the broken inversion symmetry provided by the polar EuO (111) surface.  This field mixes the spin components in the MoTe$_{2}$ monolayer, which provides overall canting of spins, especially for the edge states, as we will see later. The Rashba Hamiltonian $\mathcal{H}_{R},_{\uparrow \downarrow}$ is given by intra-site inter-$(\left|d_{xy},s \right\rangle$, $\left|d_{x^{2}-y^{2}},s \right\rangle)$ mixing terms proportional to $\lambda_{R}=72$ meV \cite{supplemental}.
All parameters are obtained from DFT calculations \cite{qi2015giant}.
Notice that this 3OTB exchange field Hamiltonian, with the right choice of TMD/substrate parameters and appropriate boundaries, could be used to study other heterostructures of interest,
such as WSe$_{2}$/CrI$_{3}$ and WS$_{2}$/MnO \cite{Zhong2017van,xu2018large}. This provides an efficient and reliable approach to study different properties and behavior of the magnetic proximity-induced magnetism \cite{2018arXiv180507942Z}. We also note the close relation between the induced MEF on the MoTe$_{2}$ ribbon and induced SOC in a graphene ribbon when on a TMD \cite{frank2018}, as the proximity in both systems can be associated with effective Zeeman fields.

\emph{Results.} We consider a zigzag ribbon with 1600 Mo-sites, with width of $\sim$125 \AA{} (40 Mo sites), and length of $\sim$144 \AA{} (40 Mo sites) along the zigzag edge (different sizes do not qualitatively change results or main conclusions here). Figure \ref{fig1}(a) shows the zigzag MoTe$_{2}$ ribbon on a EuO substrate.
The lack of inversion symmetry in the 2D MoTe$_{2}$ monolayer yields two different terminations of zigzag edges, with outer Mo or Te atoms \cite{Chu2014}, as shown in Fig.\ \ref{fig1}(b).
This asymmetry produces different edge state dispersions along the ribbon. The large intrinsic SOC in MoTe$_{2}$ competes with the proximity exchange field from the FM substrate, and leads to giant valley polarization in the 2D bulk \cite{qi2015giant}, as well as to strongly spin polarized edge-modes in the finite ribbon, as will be discussed below.

\begin{figure*}
\centering
\includegraphics[width=\linewidth]{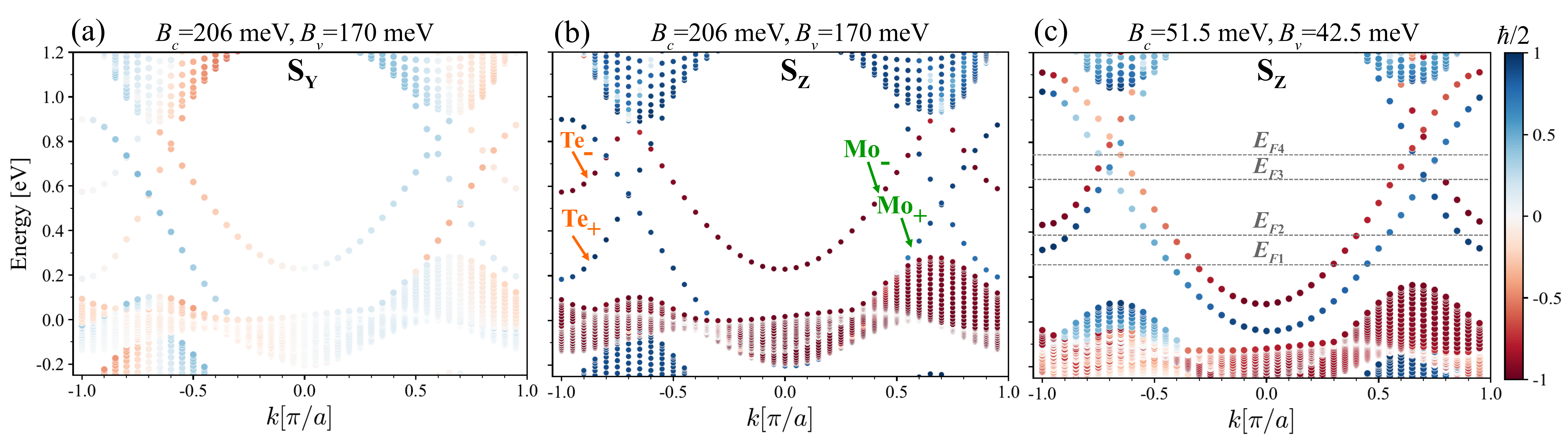}
\caption{Spin projections of energy spectrum near bulk gap along the zigzag edge of a MoTe$_{2}$ ribbon on EuO. In (a)-(b) the exchange fields for EuO $B_{v}$ and $B_{c}$ are 100\%. (a) Shows spin projections along the Y direction ($S_{\text{Y}}$); (b) along the Z direction ($S_{\text{Z}}$). In (b) Te$_{\pm}$, Mo$_{\pm}$ label in-gap 1D edge modes located on the Te and Mo-edges termination, respectively [see Fig.\ \ref{fig1} (b)]. (c) Spectrum and $S_{\text{Z}}$ for weaker exchange fields as shown; notice smaller (larger) $S_{\text{Z}}$ ($S_{\text{Y}}$) projection amplitudes than in (b) (see $S_{\text{Y}}$ in \cite{supplemental}). Gray lines indicate selected midgap Fermi levels used in Fig.\ \ref{fig3} and \ref{fig4}. Color bar indicates positive (negative) spin projection as blue (red) gradient.}
\label{fig2}
\end{figure*}

Figure \ref{fig2} shows the energy spectrum for the MoTe$_{2}$/EuO zigzag ribbon near the bulk bandgap, projected along the ribbon edge. The spectrum shows broken TRS due to exchange fields both in bulk bands, as well as on edge states dispersing through the midgap and hybridizing with bulk bands.
Panels \ref{fig2}(a)-(b) show the spin component content along $S_{\rm Y}$ and $S_{\rm Z}$, respectively, for EuO exchange fields.  For comparison, panel (c) shows the $S_{\rm Z}$ projection of the spectrum for weaker exchange fields (set to 25\% of the EuO values). Different exchange fields could be achieved by different substrate surfaces, biaxial strains, and/or van der Waals engineering of FM heterostructures \cite{Zhong2017van,li2018large}. Armchair-edge ribbons are semiconducting with gapped edge modes, as $K$ and $K$' valleys are both mapped onto the one at $\Gamma$ \cite{rostami2016edge,ridolfi2017electronic}, and producing mixing.

 For exchange fields shown, there are clear edge modes with dispersion in the bulk bandgap, and residing on either the Te-edge (labeled Te$_{\pm}$) or the Mo-edge (labeled Mo$_{\pm}$), where the subindex sign labels helicity, and states appear with significant $S_{\rm Y}$ projection, canting away from $S_{\rm Z}$, due to the Rashba coupling.
Edge modes have clear metallic behavior for Fermi levels in the bulk midgap \cite{rostami2016edge,khoeini2016peculiar}, propagating along the zigzag edges with momentum $k$ and characteristic spin \cite{Chu2014,sinova2015spin}. For the EuO full MEF, Fig.\ \ref{fig2}(a)-(b) show that the Mo$_{+}$ mode is non-degenerate and hybridized with the bulk valence band for small $k$, an effect  not present for weaker exchange fields [Fig.\ \ref{fig2}(c)] when the Mo-edge modes are fully decoupled from the bulk and located midgap.
In contrast, Te-modes are always hybridized to the bulk conduction  bands for $|k| \simeq 0.5$, regardless of the exchange field strength.
Notice the opposite group velocity of the different Mo- or Te-termination edge states at given $k$ values.

\begin{figure}
\centering
\includegraphics[width=\linewidth]{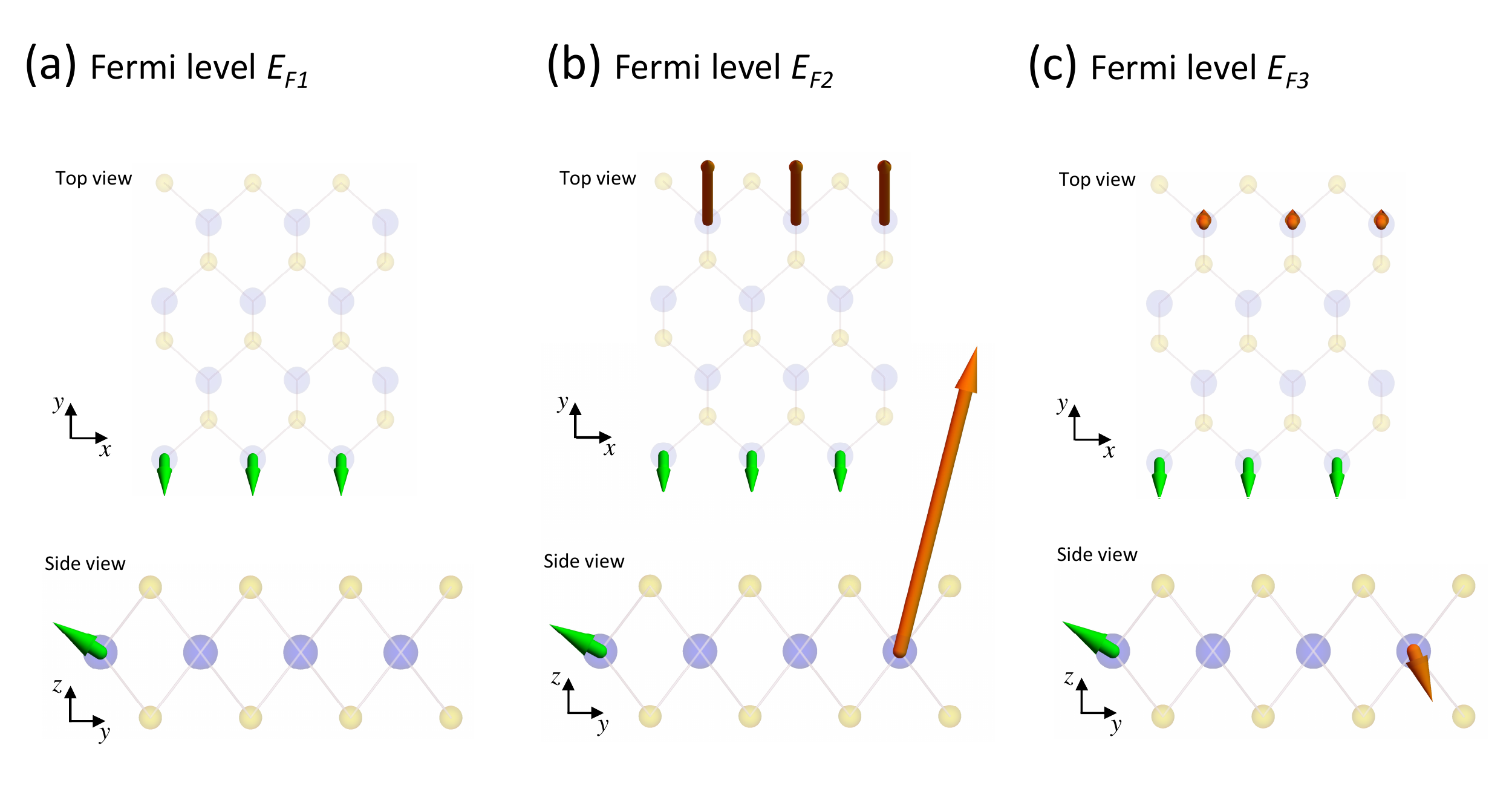}
\caption{Spin currents $\textbf{j}=(\text{j}_{\text{Y}},\text{j}_{\text{Z}})=m\textbf{j}^{\text{spin}}/\hbar$ for system in Fig.\ \ref{fig2}(c), for both Mo (green arrows) and Te (orange arrows) ribbon edges. Results for different Fermi levels, $E_{F1}$ (a), $E_{F2}$ (b) and $E_{F3}$ (c). The spin current is along the zigzag direction ($k>0$). The arrow's size (direction) indicates the magnitude (orientation) of the spin current. The magnetic substrate is not shown and the ribbon size is only schematic. $E_{F4}$ yields similar results to $E_{F3}$ (not shown).}
\label{fig3}
\end{figure}

\begin{figure}
\centering
\includegraphics[width=1.00\linewidth]{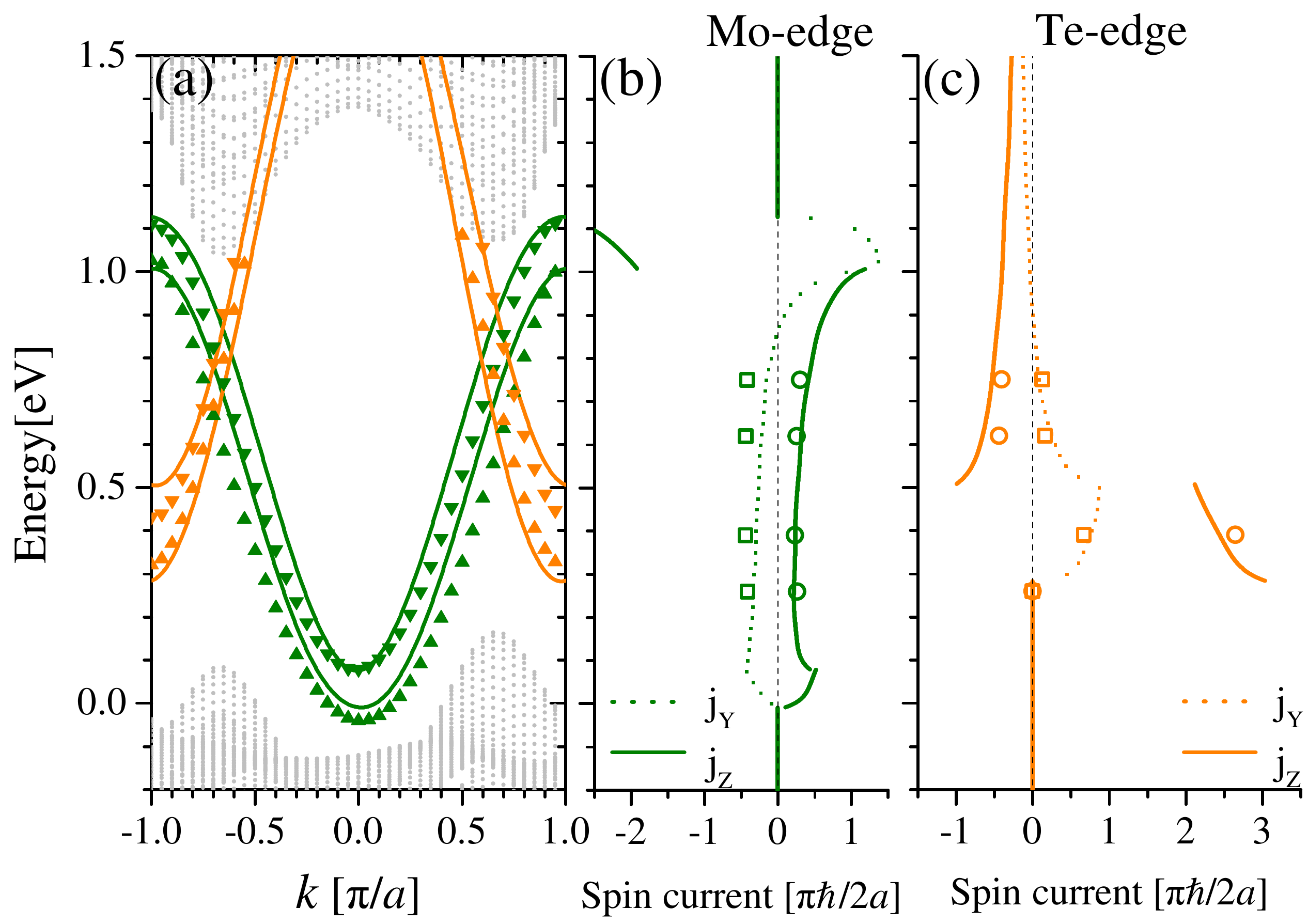}
\caption{(a) Analytical bands for Mo-edge (green lines) and Te-edge (orange lines) from Eq.\ (\ref{EffectiveHamiltonianWithPauli}) and tight-binding bands (gray dots and color triangles); Mo and Te$_{+,-}$ bands highlighted as up/down color arrows, for system in Fig.\ \ref{fig2}(c). (b) Spin current components j$_{\text{Z}}$ (solid) and j$_{\text{Y}}$ (dotted) obtained from analytical model for Mo-edge, and (c) for Te-edge, for $k>0$. Symbols in (b) and (c) are numerical values as seen in Fig.\ \ref{fig3}. The 1D effective analytical model captures the dispersions, wave functions and the spin currents of the TB results.}
\label{fig4}
\end{figure}

The EuO substrate breaks inversion symmetry, allowing a Rashba field that generates spin mixing and canted spins for the edge states \cite{ochoa2013spin}. Here, the Rashba field is along the $y$-axis, confining the spin dynamics to the YZ plane \cite{zhang2014equilibrium,sinova2004universal}.
It is clear that weaker exchange fields result in reduced $S_{\text{Z}}$ polarization, as evident in Fig.\ \ref{fig2}(b)-(c), with larger $S_{\text{Y}}$ projection as the ratio of $\lambda_R/B_v$ increases \cite{supplemental}.
As $\lambda_R$ is in principle tunable via gate fields for a given substrate-specific exchange field $B_v$, it is reasonable to anticipate that the overall spin projection (or canting) could be tunable in a given structure at specific Fermi energy.

Essential elements in spintronics include being able to inject, manipulate or detect spin polarization \cite{vzutic2004spintronics,wolf2001spintronics}. The Te- and Mo-edge modes are strongly spin-polarized along $S_{\rm Z}$--see Fig.\ \ref{fig2}.   Te$_{\pm}$ and Mo$_{\pm}$ modes with opposite momentum ($k\rightarrow -k$) propagate in opposite directions with the same $S_{\rm Z}$ projection, while the $S_{\rm Y}$ component reverses for opposite momentum, as in Fig.\ \ref{fig2}(a). This behavior is unchanged for larger $\lambda_R/B_v$ ratios, although with larger $S_{\rm Y}$ projection (see \cite{supplemental}), as the Rashba field is effectively stronger.

To characterize the propagation along the 1D Te- and Mo-edges, we select different in-gap Fermi levels to calculate the spin currents $\textbf{j}^{\text{spin}}$.  The Fermi level can be shifted by an overall gate field perpendicular to the TMD layer, allowing for tunable spin current values and polarizations \cite{lazic2016effective,dankert2017electrical}.  The spin current component at a given Fermi level is proportional to the momentum and to the spin projection, as $\text{j}_{l}^{\text{spin}}= k\,\text{S}_{l}\hbar/m$ \cite{sinova2004universal, sinova2015spin, zhang2014equilibrium},  with $l=$ Y, Z, given the corresponding spin projection, and $m(k)$ the carrier mass.
Figure \ref{fig3} shows  spin currents for the MoTe$_{2}$ ribbon in Fig.\ \ref{fig2}(c), with $\text{j}_{l}=m\text{j}_l^{\text{spin}}/\hbar= k\,\text{S}_{l}$. Given that the bulk current vanishes for in-gap Fermi levels, the non-vanishing spin currents for such levels are contributed by only the Mo- and Te-edge states and propagate along the edges of the ribbon. A 1D spin current along the Mo-edge is shown for right-movers ($k>0$) at $E_{F1}$ in Fig.\ \ref{fig3}(a). At this Fermi level, both spin-split Mo-modes contribute to the spin current with j$_{\text{Y}}$ and j$_{\text{Z}}$ pointing to negative and positive directions, respectively--notice no Te-modes contribute yet at this level. As higher Fermi levels are reached, as in the case of $E_{F2}$ and $E_{F3}$, the spin-polarized Te-modes are turned on, and contribute to the spin currents, as shown in Fig.\ \ref{fig3}(b)-(c).

The spin currents along the Mo-edge are small in magnitude, and have nearly the same polarization for all chosen Fermi levels, as shown by the green arrows in Fig.\ \ref{fig3}, as the spin projections for both spin-split Mo modes nearly cancel each other. The spin currents along the Te-edge vary drastically with Fermi level, orange arrows in Fig.\ \ref{fig3}.
The spin current for $E_{F2}$ has a large j$_{\text{Z}}$ component and non-vanishing j$_{\text{Y}}$, as only the Te$_{+}$ mode contributes.
However, the Te-edge spin current becomes small and with reverse polarization for $E_{F3}$ (or  $E_{F4}$), as both Te$_\pm$ modes contribute with nearly the same magnitude and opposite polarization.
Accordingly, one could modulate the spin-polarized currents along the Mo-edge, or simultaneously along the Te-edge of the zigzag ribbon, by tuning the Fermi level across the structure \cite{lazic2016effective,dankert2017electrical}. Similar spin-polarization in graphene nanoribbons has been proposed as spin injector device \cite{Yu2015APL,2018arXiv180507942Z}, with perhaps some practical advantages in the current TMD-based structure.

{\em Effective 1D model}.
The effective 1D Hamiltonian for the hybrid MoTe$_{2}$/EuO edges is
\begin{eqnarray}\label{EffectiveHamiltonianWithPauli}
\mathcal{H}_{\text{eff}}^{\alpha}(k) &=& \varepsilon^{\alpha} -\alpha [\hat{\sigma}_{z}+1]t^{\alpha}_{\uparrow}\cos{k}+\alpha [\hat{\sigma}_{z}-1]t^{\alpha}_{\downarrow}\cos{k} \nonumber\\
&&+\alpha \hat{\sigma}_{z}(t^{\alpha}_{SO}\sin{k} +b^{\alpha}) -\hat{\sigma}_{y}t^{\alpha}_{R}\sin{k},
\end{eqnarray}
where $\alpha$ indicates Mo ($\alpha=1$) or Te ($\alpha=-1$) edges, in terms of onsite energies $\varepsilon^{\alpha}$, effective bandwidths for the spin up/down $t^{\alpha}_{\uparrow/\downarrow}$ bands, as well as Rashba $t^{\alpha}_{R}$, and diagonal SOC $t^{\alpha}_{SO}$ and exchange fields $b^{\alpha}$.
The edge dispersion calculated from Eq.\ \ref{EffectiveHamiltonianWithPauli} is shown in Fig.\ \ref{fig4}(a). There is excellent agreement between  numerical results (symbols) and the fitted model (lines) \cite{fitting} for all Mo- and Te-modes. This Hamiltonian captures the spin content of the edge state dispersions and allows one to easily obtain the spin currents for the system.

Spin currents for Mo and Te modes, shown in Fig.\ \ref{fig4}(b)-(c), are sizable for Fermi levels in the bulk gap region.
Notice the Mo$_{+}$ mode is never singly-populated, as bulk states in the valence band are also reached before Mo$_-$ is populated, for $E_F \lesssim 0.1$ eV.\@ For $E_F \sim$0.15 to 0.3 eV, both Mo-edge modes are populated, and the spin current remains nearly constant and only present on that edge throughout that $E_F$-window.  As the Te$_{+}$ mode is reached only one spin branch at the Te-edge is populated for $E_F \sim$0.3 to 0.45 eV, with a correspondingly large spin current on that border.  The current drops when the Te$_-$ mode is reached for $E_F \gtrsim 0.5$ eV.\@ The spin current varies slowly, decreasing as the Fermi level reaches the conduction band.

Although nearly-perfect sections of both Te- and Mo-edges are found in experiments \cite{Cheng2017nanoletters,Chen2017natcomms,Zhao2017nanoletters}, we have also studied the role of structural defects on the spin currents.
On- or near-edge vacancies on either Mo- (Mo$_{\text{v}}$) or Te-sites (Te$_{\text{v}}$) produce backscattering that affects edge states and their corresponding spin current contributions. Bulk defects, in contrast, result in localized states uncoupled from the edges in realistic ribbon widths.  However, low vacancy concentrations ($\lesssim 3\%$ of on-edge defects) reduce the spin currents but only slightly, decreasing j$_{\rm Y}$ on the Mo-edge, while leaving it nearly unchanged on the Te-edge. On both edges, j$_{\rm Z}$ remains qualitatively the same for energies away from the onset of the midgap states, $E\gtrsim 0.5$ eV, as the lower density of states reduces the effect of backscattering.  As described in detail in \cite{supplemental}, the spin currents fall with increasing defect concentration, but persist over realistic ranges in experimental samples.  As flake quality is increasingly improving, we anticipate this would not limit the observability of the spin currents for midgap Fermi energies.

One should also consider the effect of electronic interactions on the spin currents, anticipating the competition with antiferromagnetic ordering across the ribbon or along the edge seen in graphene systems \cite{Jung2009,Magda2014}.  However, the presence of the gap across the ribbon, and the strong substrate-induced MEF, effectively reduce the role of interactions.  Mean-field estimates suggest that the strong exchange field bias dominates, especially away from half-filling, contributing to TRS breaking and resulting in  polarization of the edges with slightly renormalized parameters due to the Coulomb interaction \cite{supplemental}.

These results indicate that a finite size ribbon of a TMD monolayer (such as MoTe$_{2}$)  on a FM substrate (such as EuO) could be used to produce tunable spin currents along the edges of experimental samples, even for moderate vacancy concentrations (see Supplemental Material \cite{supplemental}). Such proximity-induced functionality would contribute to the rich behavior of different van der Waals systems.


\emph{Conclusions.} 
The broken inversion and time-reversal symmetries in a proximitized TMD ribbon lying on a ferromagnetic substrate splits the electronic edge states residing in the bulk midgap and produce effective 1D conducting channels with spin-polarized currents. Competition between the effective exchange and Rashba fields generates canting of the spin orientation of the spin currents. The spin current polarization and onset could be modulated by tuning the Fermi level \cite{lazic2016effective,dankert2017electrical}, and/or the effective exchange field by van der Waals engineering of heterostructures \cite{Zhong2017van}, or through biaxial strain \cite{li2018large}. The ready availability of samples and the flexibility of this effect suggests that such proximitized TMD ribbons could be effectively used as robust 1D  spin injectors \cite{2018arXiv180507942Z}. We also look forward to studies of electronic interactions in these 1D channels, involving strong spin-orbit coupling and broken symmetry.


\emph{Acknowledgments.} N.C. acknowledges support from Conicyt grant 21160844, DGIIP USM and the hospitality of Ohio University. L.R. and P.A.O. acknowledge FONDECYT grant 1180914 and DGIIP USM internal grant.  O.A.-O. and S.E.U. acknowledge support from NSF grant DMR 1508325.


\bibliography{CortesEtAl}

\end{document}